\documentclass[superscriptaddress,aps,twocolumn,longbibliography,nofootinbib]{revtex4-2}
\pdfoutput=1
\usepackage{graphicx, color, graphpap}      
\usepackage{amsmath}
\usepackage{enumerate}
\usepackage{amssymb}
\usepackage{amsthm}
\usepackage{pstricks}
\usepackage{float}
\usepackage[colorlinks=true, citecolor=blue, linkcolor=red]{hyperref}
\usepackage[T1]{fontenc}
\usepackage{bbm}
\usepackage{dsfont}
\usepackage[linesnumbered,ruled,vlined]{algorithm2e}
\SetKwInput{kwInit}{Init}
\usepackage{mathtools}

\usepackage{tikz}
\usepackage{graphicx}
\usetikzlibrary{positioning}
\usepackage{graphicx}
\usepackage{xcolor}
\usepackage[caption=false]{subfig}
\usepackage[ruled,vlined]{algorithm2e}
\usepackage{siunitx}
\usepackage{qcircuit}
\usepackage{pifont} 
\usepackage{todonotes}
\usepackage{physics}
\usepackage{thmtools}
\usepackage{thm-restate}
\usepackage{multirow}


\renewcommand{\ket}[1]{| #1 \rangle}
\renewcommand{\bra}[1]{\langle #1 |}

\renewcommand{\vec}[1]{\boldsymbol{#1}}  

\renewcommand{\eqref}[1]{(\ref{#1})}

\newtheoremstyle{example}{\topsep}{\topsep}%
{}
{}
{\bfseries}
{:}
{   }
{\thmname{#1}\thmnumber{ #2}}
\theoremstyle{example}

\theoremstyle{definition}

\newtheorem*{theorem*}{Theorem}

\def\orcid#1{\kern -0.4em\href{https://orcid.org/#1}{\includegraphics[keepaspectratio,width=0.7em]{orcid_logo.pdf}}}

\renewcommand{\>}{\rangle}
\newcommand{\<}{\langle}

\renewcommand{\H}{\mathcal{H}}

\long\def\ca#1\cb{} 
\begin{document}
\title{The cost of quantum algorithms for biochemistry:\\A case study in metaphosphate hydrolysis}

\author{Ryan LaRose} \thanks{Corresponding author:  \href{rmlarose@msu.edu}{rmlarose@msu.edu}}
\affiliation{Department of Computational Mathematics, Science, and Engineering, Michigan State University, East Lansing, MI 48823, USA}
\affiliation{Department of Electrical and Computer Engineering, Michigan State University, East Lansing, MI 48823, USA}
\affiliation{Department of Physics and Astronomy, Michigan State University, East Lansing, MI 48823, USA}
\affiliation{Center for Quantum Computing, Science, and Engineering, Michigan State University, East Lansing, MI 48823, USA}

\author{Antonios M. Alvertis}
\affiliation{Department of Physics, The University of Texas at Austin, Austin, TX 78712, USA}
\affiliation{Oden Institute for Computational Engineering and Sciences, The University of Texas at Austin, Austin, TX 78712, USA}

\author{Alan Bidart}
\affiliation{Department of Chemistry, Brown University, Providence, RI, USA, 02912}

\author{Ben DalFavero}
\affiliation{Department of Computational Mathematics, Science, and Engineering, Michigan State University, East Lansing, MI 48823, USA}
\affiliation{Center for Quantum Computing, Science, and Engineering, Michigan State University, East Lansing, MI 48823, USA}

\author{Sophia E. Economou}
\affiliation{Department of Physics, Virginia Tech, Blacksburg, VA 24061}
\affiliation{Virginia Tech Center for Quantum Information Science and Engineering, Blacksburg, VA 24061}

\author{J. Wayne Mullinax}
\affiliation{KBR, Inc., Intelligent Systems Division, NASA Ames Research Center, Moffett Field, CA 94035, USA}

\author{Mafalda Ram\^oa}
\affiliation{Department of Physics, Virginia Tech, Blacksburg, VA 24061}
\affiliation{Virginia Tech Center for Quantum Information Science and Engineering, Blacksburg, VA 24061}
\affiliation{International Iberian Nanotechnology Laboratory (INL), Portugal}
\affiliation{High-Assurance Software Laboratory (HASLab), Portugal}
\affiliation{Department of Computer Science, University of Minho, Portugal}

\author{Jeremiah Rowland}
\affiliation{Department of Physics and Astronomy, Michigan State University, East Lansing, MI 48823, USA}
\affiliation{Center for Quantum Computing, Science, and Engineering, Michigan State University, East Lansing, MI 48823, USA}

\author{Brenda Rubenstein}
\affiliation{Department of Chemistry, Brown University, Providence, RI, USA, 02912}
\affiliation{Department of Physics, Brown University, Providence, RI, USA, 02912}
\affiliation{Data Science Institute, Brown University, Providence, RI, USA, 02912}

\author{Nicolas PD Sawaya}
\affiliation{Azulene Labs, Berkeley, CA 94720}

\author{Prateek Vaish}
\affiliation{Department of Chemistry, Brown University, Providence, RI, USA, 02912}

\author{Grant M. Rotskoff}
\affiliation{Department of Chemistry, Stanford University}
\affiliation{Institute for Computational and Mathematical Engineering, Stanford University}

\author{Norm M. Tubman}
\affiliation{Applied Physics Group, NASA Ames Research Center}

\begin{abstract}
We evaluate the quantum resource requirements for ATP/metaphosphate hydrolysis, one of the most important reactions in all of biology with implications for metabolism, cellular signaling, and cancer therapeutics. In particular, we consider three algorithms for solving the ground state energy estimation problem: the variational quantum eigensolver, quantum Krylov, and quantum phase estimation. By utilizing exact classical simulation, numerical estimation, and analytical bounds, we provide a current and future outlook for using quantum computers to solve impactful biochemical and biological problems.  Our results show that variational methods, while being the most heuristic, still require substantially fewer overall resources on quantum hardware, and could feasibly address such problems on current or near-future devices. We include our complete dataset of biomolecular Hamiltonians and code as benchmarks to improve upon with future techniques.
\end{abstract}

\maketitle

\tableofcontents

\section{Introduction}
Quantum computers have the potential to significantly impact the study of important biochemical processes because atomic and molecular systems can be naturally embedded in quantum information processors. While this general fact has been appreciated for over 45 years~\cite{Feynman_1982}, today, the challenge still remains to demonstrate impactful applications of quantum technology, both in biochemistry (which we focus on in this work) as well as in other areas of chemistry and physics. While efficient quantum algorithms for simulating chemical systems are known, the primary reason these algorithms have not produced transformative results yet is because of their high overhead, both in general and in relation to the capabilities of current quantum computers in particular. While the scaling laws for quantum algorithms are comparatively easy to work out, practical overhead costs for particular problems are relatively understudied, save for a few example systems such as FeMoCo~\cite{reiher2017elucidating,lee2021even,lee2023evaluating}. Recently, resource estimation work, particularly in Trotter error for chemical Hamiltonians, has demonstrated that numerical evaluation and/or estimation can produce much tighter resource estimates than analytical bounds, giving a more accurate picture~\cite{kivlichan2020improved,gunther2025phase,chi2024average,babbush2015chemical}. Further, the end-to-end cost of quantum protocols can often be opaque from asymptotic scaling results when full details of the problem and hardware are factored in. For example, while randomized algorithms for measuring expectation values on quantum computers are asymptotically optimal~\cite{Huang_Kueng_Preskill_2020}, simpler protocols utilizing problem-specific information have been shown to require fewer resources in practice~\cite{dutt2023practical}. 

Our goal in this work is to evaluate the end-to-end resources required to solve an electronic structure problem of relevance to biochemistry. With respect to quantum resource estimation, our work extends the literature in three main ways. First, while resource estimation generally focuses on one algorithm, we provide detailed analyses for three quantum algorithms. These algorithms --- the variational quantum eigensolver, quantum Krylov, and quantum phase estimation --- are generally considered to fall under three ``eras'' of quantum computing: current noisy intermediate-scale quantum (NISQ) computing, near-future devices capable of millions of quantum operations (MegaQuop computing), and fault-tolerant, application scale quantum (FASQ)  computing~\cite{Preskill_2018,Preskill_2025}. Thus, our work provides a picture of the current and future outlook of quantum algorithms. Second, resource estimation is often general and does not take particular details about quantum computer hardware and/or software into account. However, in our work, we analyze resources both generally and compiled to particular quantum hardware platforms, utilizing both theoretical results, numerical simulation, and state-of-the-art quantum compilation methods. Third, we focus on a problem in biochemistry, magnesium-coupled metaphosphate hydrolysis, which has, to the best of our knowledge, not yet been considered in the quantum resource estimation literature. This problem is interesting and important to consider for its potential impact in biochemistry, biology, and human health. Indeed, as we elaborate on later, ATP hydrolysis is a crucial reaction with implications for metabolism, cellular signaling, and cancer therapeutics. Second, this problem is interesting to consider for quantum algorithms because, as we will show through classical pre-processing and other methods, it is feasible to treat this problem in NISQ, MegaQuop, and FASQ computing settings.

In what follows, we provide background in Sec.~\ref{sec:preliminaries}, covering ATP/metaphosphate hydrolysis as well as other problems we consider in Sec.~\ref{sec:problem-definition}. We also define the ground state energy estimation problem and describe the three quantum algorithms we consider for this task in Sec.~\ref{sec:algorithms}. Our results are presented in Sec.~\ref{sec:results} and summarized in Table~\ref{tab:resources}. Finally, we describe the methods we use to determine resource counts for each algorithm in Sec.~\ref{sec:methods}, including Hamiltonian downfolding (Sec.~\ref{sec:methods-downfolding}), the (CEO-)ADAPT method for VQE (Sec.~\ref{sec:methods-adapt})  \cite{grimsleyAdaptiveVariationalAlgorithm2019,ramoaReducingResourcesRequired2025}, quantum subroutines for time evolution (Sec.~\ref{sec:methods-time-evolution},) classical simulation methods (Sec.~\ref{sec:methods-classical-simulation}), and numerical evaluation of bounds for resources in quantum Krylov (Sec.~\ref{sec:methods-krylov}) and quantum phase estimation (Sec.~\ref{sec:methods-qpe}).

\subsection{Preliminaries} \label{sec:preliminaries}

\subsubsection{Problem definitions} \label{sec:problem-definition}

\begin{figure}
    \centering
    \includegraphics[width=\linewidth]{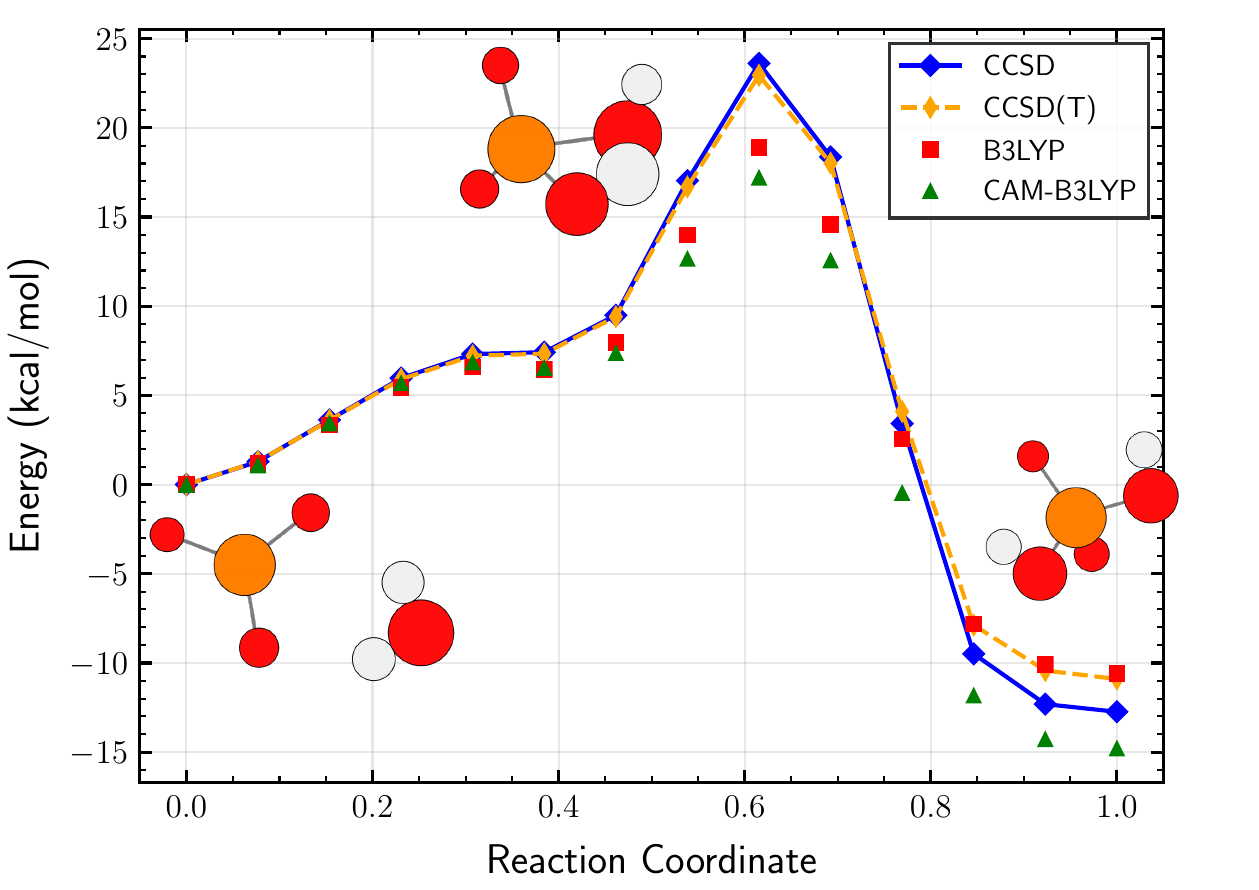}
    \caption{Energy of the metaphosphate hydrolysis reaction as a function of its reaction coordinate according to several different classical electronic structure theories. These theories differ most at the transition state, which is an indication of the quantum mechanical difficulty of the problem.}
    \label{fig:metaphoshate_reaction_path_methods}
\end{figure}

\paragraph{ATP and Metaphosphate Hydrolysis}

One of the most important reactions in all of biology is adenosine triphosphate (ATP) hydrolysis, the cleaving of the terminal phosphate bond of ATP by water to produce adenosine diphosphate (ADP) and inorganic phosphate. 
This reaction provides an essential source of irreversibility in living systems, releasing -7.3 kcal/mol of energy per phosphate cleavage, which is harnessed for metabolism, cellular signaling, and myriad other reactions important for biological function \cite{kamerlin2009energetics, westheimer1987nature}. 
Because of these essential roles in signaling and metabolism, specific inhibition of ATP binding and hydrolysis is a longstanding pharmaceutical target across the proteome~\cite{chiarella2021extracellular}. 

Despite this reaction's importance, however, determining its atomistic mechanism based on first-principles remains a grand challenge in the fields of biomolecular and \textit{ab initio} modeling. While experiments have determined the rough free energy difference between the reaction's reactants and products, many decades of computational modeling have yet to predict this free energy difference \cite{kamerlin2013nature}. This is because a correct description of the reaction not only necessitates a correct description of complicated bond breaking, which requires highly accurate quantum mechanical theories to reproduce, but also because of complex solvent effects and the structure and dynamics of the ATP molecule itself as the reaction transpires. Previous state-of-the-art studies have employed Density Functional Theory (DFT) within a Quantum Mechanics/Molecular Mechanics (QM/MM) framework, but have either significantly under- or overshot the known free energy difference and activation energy, sometimes even predicting the reaction to be endothermic overall \cite{kamerlin2009energetics, akola2003atp}. This state-of-affairs calls for the application of more accurate computational approaches, including quantum computation, to this problem. 

The full ATP hydrolysis reaction involves many dozens of atoms that would be beyond the reach of current quantum hardware. Because we are interested in resources for problems addressable by NISQ, MegaQuop, and FASQ computers, we take metaphosphate hydrolysis as our primary problem for which to estimate resources in this work. In metaphosphate hydrolysis, metaphosphate (PO$_3^{-}$) effectively lyses a water molecule, adding a hydrogen to one of its oxygens and an OH to the P atom to become (PO$_4$H$_2^{-}$)~\cite{plotnikov_quantifying_2013}. 
In a minimal, 6-31G(d) basis, not considering other solvent molecules explicitly, the reaction involves 50 electrons in 78 orbitals, making it amenable to modeling on quantum circuit simulators.

As depicted in Figure \ref{fig:metaphoshate_reaction_path_methods}, the energetics of this reaction are highly dependent on the choice of the electronic structure method. The B3LYP functional predicts an activation barrier of 18.92 kcal/mol, while CAM-B3LYP predicts 17.23 kcal/mol. In contrast, the high-accuracy coupled cluster (CCSD(T)) method yields a barrier of 22.95 kcal/mol. The reaction energies also show considerable variation: while the B3LYP value (-10.57 kcal/mol) is in close agreement with the CCSD(T) benchmark (-10.90 kcal/mol), the CAM-B3LYP functional predicts a significantly more exothermic reaction (-14.77 kcal/mol). This discrepancy between DFT and CCSD(T) suggests that higher accuracy methods are needed to accurately predict the hydrolysis reaction mechanism. 

\paragraph{Systems for scaling: Hubbard models and Hydrogen chains}

In addition to metaphosphate hydrolysis, we also consider simpler systems which we can scale, namely the spinless (Fermi-)Hubbard Hamiltonian
\begin{equation} \label{eqn:hubbard}
    \H = - t \sum_{\langle i, j \rangle} (a^\dagger_i a_j + a^\dagger_j a_i)
    + U \sum_{\langle i, j \rangle} a^\dagger_i a_i a^\dagger_j a_j
    - \mu \sum_i a_i^\dagger a_i
\end{equation}
and a linear chain of hydrogen atoms, both of which are commonly used as molecular benchmarks~\cite{huggins2021efficient}. In the Hubbard model, we consider an $n_x \times n_y$ lattice, which requires $n_x n_y$ qubits, and for the hydrogen chain, we consider $n_H$ atoms, which require $2 n_H$ qubits. For both systems, we map the fermionic Hamiltonians to qubit Hamiltonians via the Jordan-Wigner transformation.

\subsubsection{Quantum algorithms for ground state energy estimation} \label{sec:algorithms}

We consider the following quantum algorithms for the task of ground state energy estimation --- that is, given a Hamiltonian $\H$ and desired accuracy $\epsilon$, output an estimate $E$ of the ground state energy $E_0$ such that 
\begin{equation} \label{eqn:ground-state-energy-estimation-problem}
    |E - E_0 | \le \epsilon .
\end{equation}
Throughout the paper, we work in units of Hartree and target chemical accuracy, taken to be $\epsilon = 10^{-3}$ Ha.

\paragraph{Variational quantum algorithms}
\label{sec:VQAs}

The variational quantum eigensolver (VQE) was introduced with the aim of reducing the resources for quantum algorithms, especially on early quantum computers~\cite{peruzzo2014variational}. In VQE, a trial wavefunction (``ansatz'') $|\psi\rangle$ is parameterized $|\psi\rangle = |\psi(\theta)\rangle$ and the energy is minimized to produce
\begin{equation}
    E := \min_{\theta} \langle \psi(\theta) | H | \psi(\theta) \rangle 
\end{equation}
as the ground state energy estimate. In other words, VQE is simply the well-known variational principle where the trial wavefunction is prepared on a quantum computer. Shortly after VQE, another variational quantum algorithm known as the QAOA was introduced~\cite{farhi2014quantum}, and since then, the area has attracted significant attention~\cite{cerezo2021variational}. The primary advantage of variational algorithms is that they limit the quantum resources required relative to non-variational quantum algorithms. The primary disadvantage of variational algorithms is that they have limited or no guarantees on runtime or convergence.

\paragraph{Quantum Krylov}

Quantum Krylov algorithms utilize the well-known Krylov method in which, given an operator $A$ and a vector $|b\rangle$, the $d$-dimensional \textit{Krylov subspace}
\begin{equation}
    \mathcal{K}_A := \{ |\phi_j\rangle \}_{j = 0}^{d - 1} := \{ A^j |b\rangle \}_{j = 0}^{d - 1}
\end{equation}
is formed, then the generalized eigenvalue problem
\begin{equation} \label{eqn:krylov-generalized-eigenvalue-problem}
    H \vec{c} = E S \vec{c}
\end{equation}
is solved to output the ground state energy estimate $E$~\cite{epperly2022theory}, where the matrices $H$ and $S$ are defined as
\begin{equation} \label{eqn:krylov-matrix-elements}
    H_{ij} := \< \phi_i| A | \phi_j\> \qquad \text{ and } \qquad S_{ij} := \< \phi_i| \phi_j \> . 
\end{equation}
In quantum Krylov, a quantum computer is used to compute the matrix elements~\eqref{eqn:krylov-matrix-elements}, then the generalized eigenvalue problem is solved classically. To compute matrix elements on a quantum computer, the Hadamard test is typically used~\cite{yoshioka2025diagonalization}. The Hadamard test is a standard quantum subroutine~\cite{nielsen2010quantum} which returns matrix elements of the form $\< \phi_i | U | \phi_j \>$ provided circuits for preparing $|\phi_i\rangle$, $|\phi_j\>$, and applying the unitary $U$. The subroutine requires one ancilla qubit and implements controlled versions of the preparation circuits/unitaries. Because the Hamiltonian $\H$ is typically not unitary, one generally implements quantum Krylov with the unitary time evolution operator $A_t := \exp(-i \H t)$ for time parameter $t$ when creating the Krylov subspace vectors $|\phi_j\>$. (Note that when evaluating matrix elements~\eqref{eqn:krylov-matrix-elements}, the Hamiltonian $\H$ is still used.) In this case, powers of $A_t$ have the form $A_t^j = A_{jt} = \exp( -i \H t j)$, corresponding to longer time evolution on the quantum processor. It is known that (classical) Krylov methods utilizing $A = \H$ converge exponentially to the ground state energy in the subspace dimension $d$. Some analogous results have been demonstrated for quantum Krylov methods with $A = \exp(-i \mathcal{H} t)$~\cite{yoshioka2025diagonalization,kirby2024analysis}.

\paragraph{Quantum phase estimation}

Quantum phase estimation (QPE) is a textbook quantum algorithm~\cite{nielsen2010quantum} for computing eigenvalues of an $n$-qubit unitary operator $U$. As with quantum Krylov, the time evolution operator $U := \exp(-i \H t)$ for time parameter $t$ is nominally used for the ground state energy estimation problem. Because the operator $U$ is unitary, eigenvalues can be written in the form $e^{2 \pi i \phi}$ where $0 \le \phi < 1$ is known as the phase. In the standard textbook form, we can compute $\phi$ to $k$ binary decimal points of accuracy by using $k$ ancilla qubits, $k$ controlled-$U^{2^t}$ operations for $t = 0, ..., k - 1$, and a $k$-qubit quantum Fourier transform. As long as $U$ is efficiently implementable --- i.e., there exists a $\text{poly}(n)$ circuit implementing $U$ --- the overall procedure is asymptotically efficient, but has high overhead due to the (controlled)-$U^{2^t}$ operations. Because of this, it is still active research to reduce the resource requirements of phase estimation. A simple modification known as iterative quantum phase estimation~\cite{dobsicek2007arbitrary} requires only one ancilla qubit at the cost of $k$ total circuits to be run. Recently, a zero-ancilla version of quantum phase estimation was introduced utilizing signal processing algorithms for phase retrieval in classical postprocessing~\cite{clinton2024quantum, russoEvaluatingEnergyDifferences2021, chanAlgorithmicShadowSpectroscopy2025}. While this version of QPE requires, to our knowledge, the least overhead (in terms of ancilla qubits and therefore likely in gates because time evolution does not need to be controlled), there is not a clear understanding of convergence. Therefore, we consider single-ancilla QPE, the convergence of which has been well-studied (see e.g.~\cite{gunther2025phase} and the references therein) as an upper bound when evaluating resource requirements for the biochemical systems in this work. Sec.~\ref{sec:methods-qpe} contains more details on the methods we use to evaluate resource requirements for QPE.

\section{Results} \label{sec:results}

\begin{table*}
    \centering
    \renewcommand{\arraystretch}{1.3}
    \begin{tabular}{|c|c|c|c|ccc|c|ccc|ccc|}
    \hline
    \multirow{2}{*}{\textbf{System}} &
    \multirow{2}{*}{$n$} &
    \multirow{2}{*}{$N_{\text{terms}}$} &
    \multirow{2}{*}{$N_{\text{groups}}$} &
    \multicolumn{3}{c|}{\textbf{ADAPT-VQE}} &
    \multirow{2}{*}{{Trotter}} &
    \multicolumn{3}{c|}{\textbf{Q. Krylov}} &
    \multicolumn{3}{c|}{\textbf{QPE}} \\ \cline{5-7} \cline{9-14}
     &  &  &  & $n_Q$ & $n_C$ & $n_{2Q}$ & $n_{2Q}$ & $n_Q$ & $n_C$* & $n_{2Q}$* & $n_Q$ & $n_C$ & $n_{2Q}^\dagger$ \\ \hline
    $\H_1 = $ PO$_3^-\cdots$H$_2$O & 44  & 575,751  & 2,737 & $44$  & 2,737 & 6,508  & $1.3 \cdot 10^7$ & $ 45$ & $5.7 \cdot 10^{7}$ & $4.5 \cdot 10^9$ & $45$  & $10$  & $2.0 \cdot 10^{16}$ \\ \hline
    $\H_2 = $ H$_2$PO$_4^-$ & 44 & 579,542  & 2,692  & $44$ & 2,692  & 7,200  & $1.1 \cdot 10^7$  & $45$ & $5.6 \cdot 10^7$ & $3.9 \cdot 10^9$ & $45$  & $10$ & $1.8 \cdot 10^{16}$ \\ \hline
\end{tabular}
    \caption{Quantum resources required for each algorithm to compute the ground state energy of the metaphosphate systems to chemical accuracy. For each system, the number of qubits $n$ and number of (Pauli) terms $N_\text{terms}$ in the Hamiltonian are shown, as well as the number of groups $N_\text{groups}$ after sorting Pauli strings into commuting sets. For each algorithm, we show the number of qubits $n_Q$, number of circuits $n_C$, and maximum number of two-qubit gates $n_{2Q}$ over all circuits. Resources with a * are estimated lower bounds and resources with a $\dagger$ are estimated upper bounds. Additional details and notes are provided in the main text.}
    \label{tab:resources}
\end{table*}

Resource counts for metaphosphate hydrolysis with the three quantum algorithms we consider are shown in Table~\ref{tab:resources}. This table shows the number of qubits $n$ and number of Pauli terms $N_\text{terms}$ after obtaining the Hamiltonian via the methods described in Sec.~\ref{sec:methods-downfolding}. Given these Hamiltonians, we sort terms into $N_\text{groups}$ commuting groups, discussed in Sec.~\ref{sec:methods-energy-estimation-subroutines}. 

Next, for each algorithm we show the required number of qubits $n_Q$, number of circuits $n_C$, and number of two-qubit gates $n_Q$, here assuming all-to-all connectivity. 
For ADAPT \cite{grimsleyAdaptiveVariationalAlgorithm2019}, $n_Q = n$, and $n_C = N_\text{terms}$ via subroutines for direct energy estimation described in Sec.~\ref{sec:methods-energy-estimation-subroutines}. Briefly, this means that the expectation value of each group is measured by appending a diagonalization circuit to the ansatz, then measuring in the computational basis. The number of two-qubit gates $n_{2Q}$ is found by explicit construction --- i.e., optimizing the ansatz via CEO-ADAPT \cite{ramoaReducingResourcesRequired2025} as described in Sec.~\ref{sec:methods-adapt} until chemical accuracy is reached. 

As both quantum Krylov and quantum phase estimation require time evolution as a subroutine, we compute the number of gates required for a single Trotter step of each system using the Paulihedral method described in Sec.~\ref{sec:methods-time-evolution}. For $\H_1$, a single Trotter step requires 6050622 single-qubit gates and 13016100 two-qubit gates. For $\H_2$, a single Trotter step requires 5351074 single-qubit gates and 11372404 two-qubit gates. As mentioned above, this assumes all-to-all connectivity --- we illustrate the overhead cost for the heavy hex architecture in Sec.~\ref{sec:methods-compilation}. The number of two-qubit gates, which are here CNOTs, are shown in Table~\ref{tab:resources}.

For quantum Krylov, we require one ancilla qubit to compute matrix elements via the Hadamard test. The number of circuits required is $n_C = d^2 N_\text{groups}$ where $d$ is the Krylov subspace dimension. (Note that the matrices in~\eqref{eqn:krylov-matrix-elements} are both hermitian.) We estimate $d \gtrsim 144$ for both systems (see Sec.~\ref{sec:methods-krylov}) to obtain the total number of circuits $n_C$ is Table~\ref{tab:resources}. Last, we estimate the required number of Trotter steps (see Sec.~\ref{sec:methods-krylov}) to be $n_T \gtrsim 50$. The largest circuit required for quantum Krylov thus consists of $n_T$ {controlled} Trotter steps. To approximate the number of two-qubit gates for a controlled Trotter step from a Trotter step with $n_{1QT}$ single-qubit gates and $n_{2QT}$ two-qubit gates, we use the fact that the controlled version of a product of unitaries is the product of controlled unitaries. This implies that the controlled Trotter step requires at most $3 n_{1QT}$ CNOTs (utilizing the fact that any controlled gate can be implemented with at most three CNOTs~\cite{nielsen2010quantum}) and at most $6 n_{2QT}$ CNOTs (utilizing the fact that a CCNOT (Toffoli) gate can be implemented with at most six CNOTs~\cite{nielsen2010quantum}). 
Thus our estimated lower bound on the number of two-qubit gates for quantum Krylov is $n_{2Q} \sim n_T (3 n_{1QT} + 6 n_{2QT}) \gtrsim 10^9$ two-qubit gates as shown in Table~\ref{tab:resources}.

For QPE, we also require one ancilla qubit to control the time evolution operator. Since the accuracy of single-ancilla iterative QPE doubles with every (binary) decimal, corresponding to one circuit, reaching chemical accuracy $\epsilon = 10^{-3} \sim 2^{-10}$ requires $n_C \sim 10$ circuits. We give an estimated upper bound on the number of two-qubit gates required for QPE by giving an upper bound on the number of Trotter steps $n_T$, compiling a single Trotter step into $n_{1QT}$ single-qubit gates and $n_{2QT}$ two-qubit gates, and estimating the cost of a controlled Trotter step as above via $n_{2Q} \sim n_T (3 n_{1QT} + 6 n_{2QT})$. By using a perturbation theory analysis of QPE and bounding its cost in Sec.~\ref{sec:methods-qpe}, we estimate $n_T \lesssim 2.2 \cdot 10^{8}$ for $\H_1$ and $n_T \lesssim 2.3 \cdot 10^8$ for $\H_2$. Using the Trotter step counts results in the values of $n_{2Q} \lesssim 10^{16}$ in Table~\ref{tab:resources}. We remark that these resource counts are comparable to recent estimates (e.g. Fig. 1 and Fig. 2 of~\cite{gunther2025phase}) for molecular Hamiltonians.

\begin{figure}
    \centering
    \includegraphics[width=\linewidth]{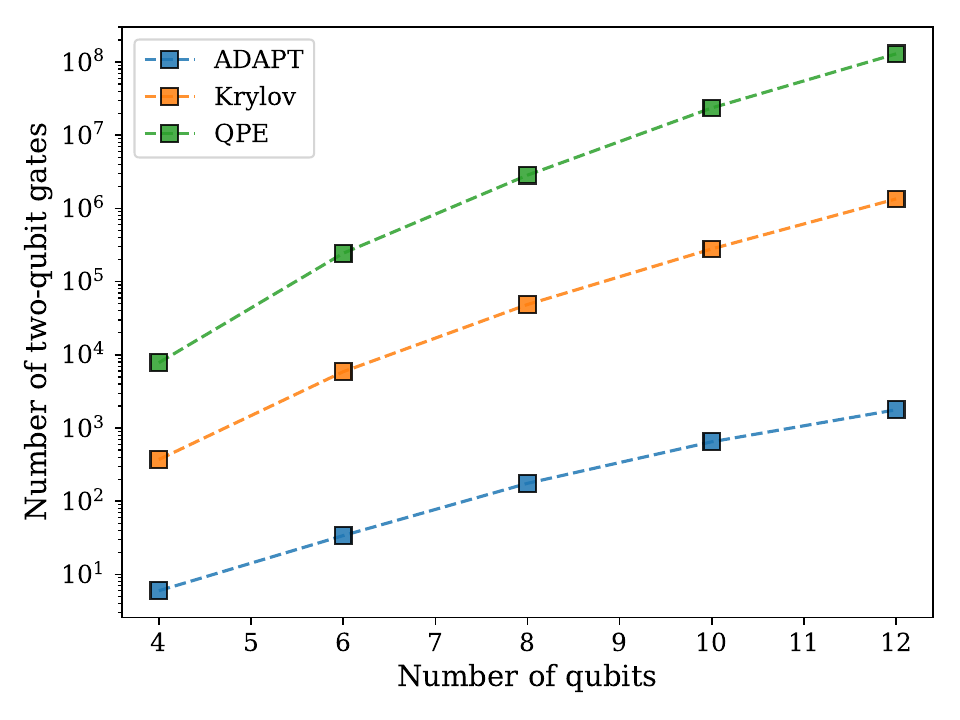}
    \caption{Number of two-qubit gates required to reach chemical accuracy for a Hydrogen chain with $n_\text{H} = n_Q / 2$ atoms. ADAPT gate counts are exact and Krylov/QPE gate counts are upper bounds.}
    \label{fig:hchain-resources}
\end{figure}

In addition to the metaphosphate systems, we analyze resources required for the same ground state energy problem when scaling a linear chain of Hydrogen atoms. Fig.~\ref{fig:hchain-resources} shows the results for $2 \le n_H \le 6$ atoms, requiring $n_Q = 2 n_H$ qubits. Here, the ADAPT resources are again calculated exactly by explicitly optimizing the circuit. The quantum Krylov circuits are also simulated exactly with varying subspace dimensions $d$ and numbers of Trotter steps $n_T$ (see Sec.~\ref{sec:methods-krylov}) to obtain the required number of two-qubit gates. In the case that the simulation achieved chemical accuracy with a given $(d, n_T)$, the results are exact. If the simulation did not achieve chemical accuracy with the largest chosen $(d, n_T)$, we estimate the required values by extrapolation. The resources for QPE use the same upper bound mentioned above and described in Sec.~\ref{sec:methods-qpe}. In addition to illustrating the scaling of each algorithm, these results agree with our general ordering of algorithmic resources in Table~\ref{tab:resources}. 

Several remarks about our results are in order. First, it is encouraging to see that ADAPT resources are well within the resources used in recent experiments, e.g.~\cite{farrell2024quantum,robledo2025chemistry}, meaning that important biological questions are within reach of being explored on current and near-future quantum computers. It is also important to note that variational quantum algorithms are designed to offload as much computation to classical processors as possible, and here we are only reporting the quantum resources. In practice, using the methods described in Sec.~\ref{sec:methods-downfolding}, Sec.~\ref{sec:methods-adapt}, and Sec.~\ref{sec:methods-classical-simulation}, we are able to synthesize quantum circuits relatively quickly for systems of these sizes, and even beyond, generally taking hours to days to perform the optimizations with state-of-the-art methods for fermionic Hamiltonians. The general exponential scaling of this classical optimization portion of variational quantum algorithms is worth noting, however, even though many ideas and workarounds have been proposed~\cite{Khatri_LaRose_Poremba_Cincio_Sornborger_Coles_2019,cerezo2021variational}. Second, we remark that the number of gates per Trotter, stemming from the complexity of the Hamiltonians we consider, is a primary bottleneck for quantum Krylov and for QPE. While we considered the lowest overhead time evolution algorithm (first order Trotter) as well as one of the best, exact compilers for Trotter circuits described in Sec.~\ref{sec:methods-time-evolution}, the number of two-qubit gates is well-beyond the capabilities of current quantum computers. However, with algorithmic improvements it is foreseeable to achieve $n_{2Q} \sim 10^6$, making it suitable for MegaQuop computers. It is likely that algorithmic improvements and/or improved measurement reduction subroutines could reduce the $n_C$ required for quantum Krylov, the bulk of which comes from the complexity of the Hamiltonians ($N_\text{terms}$ and correspondingly $N_\textrm{groups}$) and the quadratic scaling with the subspace dimension $d$. Finally, we remark that reducing the number of Trotter steps required for QPE can drastically reduce its required $n_{2Q}$.
We state these resources without regards to error correction, though it is certain that circuits at this scale will require error correction and so additional overhead.

\newpage

\section{Methods} \label{sec:methods}

\subsection{Hamiltonian downfolding} \label{sec:methods-downfolding}

Hamiltonian downfolding is a hybrid quantum-classical framework that utilizes classical computing to solve for the external orbital space and then ``folds'' this information into a new, effective Hamiltonian ($H^{\textrm{eff}}$) that operates only within a small active space. This allows the quantum algorithms to recover a significant portion of the dynamical correlation, yielding high accuracy without increasing quantum resource requirements. We apply this method to each Hamiltonian as a ``pre-processing'' step to simplify the input as much as possible uniformly across each quantum algorithm we consider.

We utilize the Double Unitary Coupled Cluster (DUCC) downfolding technique that produces a Hermitian effective Hamiltonian, making it suitable for quantum algorithms~\cite{bauman_downfolding_2019, singh2025qubit}. The DUCC ansatz partitions the cluster operator into internal ($\hat{\sigma}^{\textrm{int}}$) and external ($\hat{\sigma}^{\textrm{ext}}$) components:
\begin{equation}
    |\Psi \rangle = e^{\hat{\sigma}^{\textrm{ext}}} e^{\hat{\sigma}^{\textrm{int}}} |\phi_0 \rangle
\end{equation}

Here, $\hat{\sigma}^{\textrm{int}}$ generates excitations purely within the active space, while $\hat{\sigma}^{\textrm{ext}}$ generates all excitations involving at least one inactive orbital, and $|\phi_0 \rangle$ is the reference Hartree Fock state. This factorization allows the reduction of a full-system problem into a smaller eigenvalue problem confined to the active space:
\begin{equation}
    \hat{H}^{\textrm{eff}} e^{\hat{\sigma}^{\textrm{int}}} |\phi_0 \rangle = E e^{\hat{\sigma}^{\textrm{int}}} |\phi_0 \rangle
\end{equation}

The effective Hamiltonian $\hat{H}^{\textrm{eff}}$ is constructed by performing a similarity transformation of the full Hamiltonian $\hat{H}$ with the external operator, and then projecting the result onto the active space defined by the projector $\hat{P} + \hat{Q}_{int}$: 
\begin{equation}
    \hat{H}^{\textrm{eff}}=\left(\hat{P}+\hat{Q}_{\mathrm{int}}\right) e^{-\hat{\sigma}_{\mathrm{ext}}} \hat{H}_{\mathrm{ext}} e^{\hat{\sigma}_{\mathrm{ext}}}\left(\hat{P}+\hat{Q}_{\mathrm{int}}\right)
\end{equation}

In practice, constructing $\hat{H}^{\textrm{eff}}$ is intractable, and several approximations are required. First, the similarity transformation is a non-terminating Baker-Campbell-Hausdorff (BCH) expansion, which must be truncated at a finite order. We utilize the DUCC(3) approximation: 
\begin{equation}
    \begin{aligned} 
        \hat{H}_{\mathrm{ext}}^{\textrm{DUCC(3)}}= &\ \hat{H}_{N}+\left[\hat{H}_{N}, \hat{\sigma}_{\mathrm{ext}}\right]+\frac{1}{2}\left[\left[\hat{H}_{N}, \hat{\sigma}_{\mathrm{ext}}\right], \hat{\sigma}_{\mathrm{ext}}\right] \\ & +\frac{1}{6}\left[\left[\left[\hat{F}_{N}, \hat{\sigma}_{\mathrm{ext}}\right], \hat{\sigma}_{\mathrm{ext}}\right], \hat{\sigma}_{\mathrm{ext}}\right]
    \end{aligned}
\end{equation}
which is correct through the third order perturbation theory. Here, $\hat{H}_{N}$ and $\hat{F}_N$ are normal-order full Hamiltonian and the corresponding one-body contribution, respectively. Second, the exact $\hat{\sigma}_{\mathrm{ext}}$ is unknown and is approximated using external cluster amplitudes ($T_{\mathrm{ext}}$) obtained from a classical, full-system CCSD calculation, via the relation $\hat{\sigma}_{\mathrm{ext}} \approx \hat{T}_{\mathrm{ext}}-\hat{T}_{\mathrm{ext}}^{\dagger}$. Finally, the resulting Hamiltonian $\hat{H}^{\textrm{eff}}$, which contains many-body interactions (e.g., three- and four-body terms), is typically truncated to retain only one- and two-body terms. These calculations to obtain the one- and two-body integrals of the effective Hamiltonian were performed using the DUCC implementation within the ExaChem package~\cite{panyala_exachemexachem_2023}. The resulting effective one- and two-body integrals, which now encapsulate the external dynamical correlation effects, are used as the input for the quantum algorithms.

\subsection{ADAPT-VQE} \label{sec:methods-adapt}

The adaptive derivative-assembled problem-tailored (ADAPT)-VQE \cite{grimsleyAdaptiveVariationalAlgorithm2019} is a particular variant of VQE (as described in Sec.~\ref{sec:VQAs}) where the variational state is dynamically generated and has the form

\begin{equation}\ket{\psi{(\theta)}} = e^{\theta_N\hat{A}_N}...e^{\theta_{i}\hat{A}_{i}}...e^{\theta_1\hat{A}_1} \ket{\psi_0},
\label{eq:adapt_ansatz}
\end{equation}
with the selection of each generator $\hat{A}_i$ from a pool of candidates being informed by measurements performed on the quantum computer during the execution of the algorithm.

ADAPT-VQE starts with a classical reference state $\ket{\psi_0}$ (typically the Hartree-Fock state) and appends operators to the ansatz iteratively. At iteration $n$, the selected operator is the one that --- when appended to the current ansatz --- maximizes the derivative of the energy with respect to the variational parameter at point $\theta_n=0$. This derivative, typically referred to as `the gradient' of an operator, can be written as the expectation value of the commutator of the Hamiltonian with the candidate generator at the previous iteration's final state:

\begin{align}
\begin{split}
&\frac{\partial \bra{\psi^{(n-1)}}e^{-\theta_k\hat{A}_k}\mathcal{\hat{H}}e^{\theta_k\hat{A}_k}\ket{\psi^{(n-1)}}} {\partial\theta_k}\Bigr|_{\theta_k = 0} \\
&=\bra{\psi^{(n-1)}}[\mathcal{\hat{H}},\hat{A}_k]\ket{\psi^{(n-1)}},
\end{split}
\label{eq:adapt_commutator}
\end{align}
which provides a method for measuring these gradients on quantum computers. After the addition of each operator, a full optimization of the parameters is performed. The parameters from the previous iteration (for the operators that are already present in the circuit) serve as a warm start for the following optimization round of ADAPT.

While the original proposal was to use fermionic excitations \cite{grimsleyAdaptiveVariationalAlgorithm2019}, the generators $\hat{A}_i$ in Eq.~\ref{eq:adapt_ansatz} can take any form. Recent work on the topic of operator pools has led to significant improvements on the gate counts required to reach a given precision for a variety of systems. The qubit \cite{tangQubitADAPTVQEAdaptiveAlgorithm2021}, qubit excitation (QE) \cite{yordanovQubitexcitationbasedAdaptiveVariational2021} and coupled exchange operator (CEO) \cite{ramoaReducingResourcesRequired2025} pools all offer significant reductions in gate counts, with the latter being the most efficient pool to date. In parallel, the TETRIS variant of ADAPT-VQE \cite{anastasiouTETRISADAPTVQEAdaptiveAlgorithm2022} was proposed to decrease the ansatz depth by adding several operators acting on disjoint sets of qubits in each iteration. In this work, we employ a co-designed variant of the algorithm that produces an ansatz tailored to the target hardware \cite{co-adapt}. 

The measurements required to measure all commutators of the form of Eq.~\eqref{eq:adapt_commutator} at each operator selection step have been pointed out as a shortcoming of ADAPT-VQE. However, due to clever operator grouping strategies based on the structure of the pool operators \cite{anastasiouHowReallyMeasure2023b}, novel optimization methods \cite{jagerFastGradientfreeOptimization2025,ramoaReducingMeasurementCosts2024a}, and the naturally favorable optimization landscape enjoyed by the algorithm \cite{grimsleyAdaptiveProblemtailoredVariational2023}, ADAPT-VQE has been shown to require orders of magnitude \textit{fewer} measurements than static alternatives to reach the same precision \cite{ramoaReducingResourcesRequired2025}. Further, this algorithm has been shown to be more resilient to hardware \cite{daltonQuantifyingEffectGate2024} as well as sampling \cite{ramoaAnsatzeNoisyVariational2022} noise.

While the original proposal targeted molecular ground state preparation, generalizations of the algorithm have been proposed for a wide range of applications, including combinatorial optimization \cite{zhuAdaptiveQuantumApproximate2022}, time-periodic Hamiltonians \cite{kumarFloquetADAPTVQEQuantumAlgorithm2025}, lattice systems \cite{dykeScalingAdaptiveQuantum2023}, and preparation of mixed and excited states \cite{sambasivamTEPIDADAPTAdaptiveVariational2025,warrenAdaptiveVariationalAlgorithms2022,grimsleyChallengingExcitedStates2025,yordanovMolecularExcitedState2021a}.

To obtain the resource counts for CEO-ADAPT, we use effective Hamiltonians generated with the DUCC downfolding formalism on the starting and end points of the of the Reaction Coordinate in Fig.~\ref{fig:metaphoshate_reaction_path_methods}. The targeted active space for the downfolding was a 32 electron in 22 orbital active space, which is equivalent to a 44-qubit problem using the standard Jordan-Wigner transformation.  We employed two features of the SWCS to be able to perform ADAPT-VQE optimization for this system. First, we constrained the number of determinants when evaluating the action of the ansatz on the Hartree-Fock reference state to at most 2,000 determinants. Second, we employ a random element to the operator selection to the operator selection step in the standard ADAPT-VQE algorithm. Because the number of operators in the pool was 409,374, a number too large for our current implementation, we selected a subset of these operators at each iteration to consider when picking the new operator to add to the ansatz. The generation of this subset involves keeping 500 operators with the largest gradients from the previous ADAPT iteration and adding 500 operators randomly from the overall pool. This allows us to keep the computational cost manageable while retaining the most important operators in the ansatz.

\subsection{Quantum subroutine for time evolution} \label{sec:methods-time-evolution}

Both quantum Krylov and quantum phase estimation nominally require a time evolution subroutine for $U := \exp(-i \H t)$. Of the plethora of quantum algorithms for time evolution ranging from Taylor series~\cite{berry2015simulating} to linear combinations of unitaries~\cite{childs2010hamiltonian} to randomized methods~\cite{campbell2019random} to variational methods~\cite{benedetti_hardware-efficient_2021}, (Lie-Suzuki-)Trotter methods remain a standard method (especially in recent large-scale quantum computations~\cite{kim2023evidence,farrell2024quantum,piccinelli2025quantum}) due to (i) their ability to produce relatively short-depth circuits and (ii) the ability to arbitrarily decrease the error by increasing the number of Trotter steps.

In its simplest form, the (first order) Trotter formula reads
\begin{equation}
    e^{-i (\H_A + \H_B) t} = \left[ e^{-i \H_A t / r} e^{- i \H_B t / r} \right]^r + O(t^2 / r), 
\end{equation}
where $r = 1, 2, 3, ...$ is the number of Trotter steps. For Hamiltonians with more than two terms, the error can be upper bounded as a sum of nested commutators~\cite{childs2021theory} in addition to the $t^2 / r$ dependence. We remark that, while such analytical bounds are known, they are often loose bounds and so direct numerical evaluation is preferable. Additionally, for algorithms using time evolution as a subroutine, it is not always clear how the Trotter error propagates through to the final energy error. For these reasons, we determine costs by, whenever possible, simulating algorithms to determine the (minimum) number of Trotter steps needed and directly composing Trotter circuits to count required resources.

\begin{figure}
    \centering
    \includegraphics[width=\linewidth]{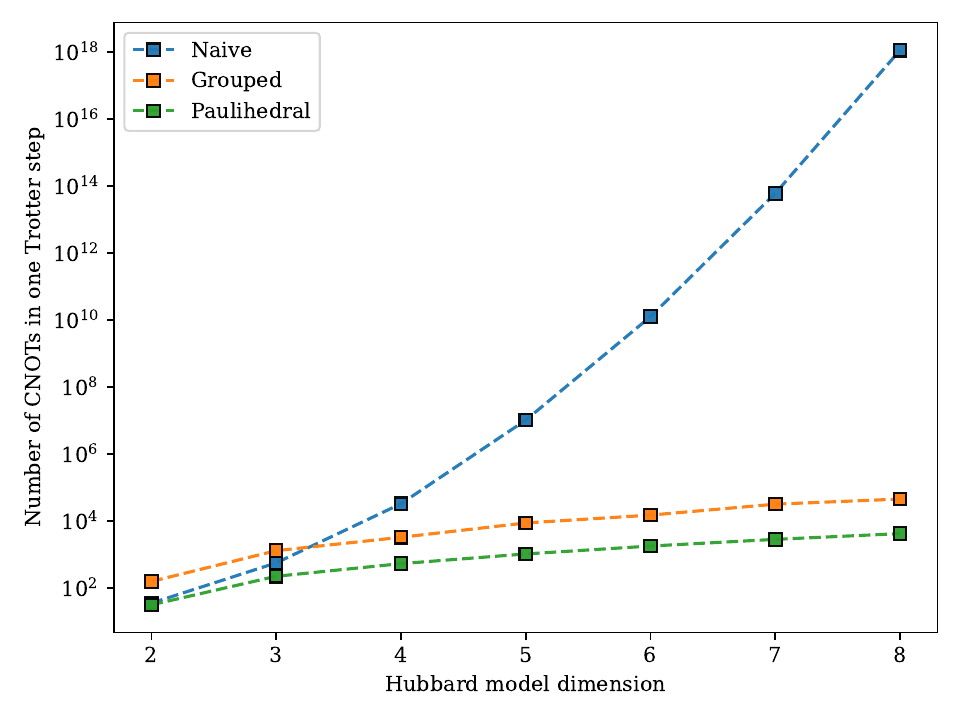}
    \caption{Number of CNOTs required for one Trotter step with three methods, the ``naive'' strategy directly exponentiating each Pauli string in the Hamiltonian, the ``grouped'' strategy where terms are first sorted into commuting sets and then exponentiated, and the Paulihedral strategy~\cite{li2022paulihedral}. All strategies consider a gateset of $\{U, \text{CNOT}\}$ where $U$ is an arbitrary single-qubit gate and CNOT can be performed between any pair of qubits. The Hamiltonian here is an $n \times n$ spinless Fermi-Hubbard model (which acts on $n^2$ qubits). While the naive and grouped strategies are upper bounds, we see the Paulihedral method produces circuits with significantly fewer two-qubit gates as the problem is scaled. For this reason we use the Paulihedral method when evaluating overall resources for quantum Krylov and quantum phase estimation (which require time evolution as a subroutine).}
    \label{fig:trotter-methods-performance}
\end{figure}

To that end, we have considered three methods for time evolution by first order Trotterization. The first is a ``naive'' method which directly exponentiates each term in order, the second groups terms into ($k$-)commuting sets~\cite{dalFavero2025measurement} and then exponentiates each group, and the third is a recent quantum simulation compiler known as Pauliedral which implements various optimizations such as gate cancellation and qubit mapping and has shown to perform as good or better than state-of-the-art compilers on a variety of benchmark problems~\cite{li2022paulihedral}. We evaluated each method on our problems, for example we show the performance of each when compiling $n \times n$ Hubbard models to the gateset $\{U, \text{CNOT}\}$ (all-to-all connectivity) in Fig.~\ref{fig:trotter-methods-performance}. From this example, as well as others we have tried, we see that Paulihedral produces Trotter circuits with the fewest two-qubit gates (as well as fewest overall gates). For this reason we use Paulihedral as the method for time evolution when evaluating overall resource counts in quantum Krylov and quantum phase estimation.

\subsection{Classical simulation} \label{sec:methods-classical-simulation}

Whenever possible, we simulate circuits directly to obtain exact resource counts, for example in Fig.~\ref{fig:hchain-resources} for ADAPT and quantum Krylov. We also utilize classical simulations to estimate algorithm parameters necessary to achieve chemical accuracy, for example the subspace dimension $d$ in quantum Krylov. An example of this is shown in Fig.~\ref{fig:krylov-convergence-hchain} where we numerically determine the minimum subspace dimension and number of Trotter steps to converge for a Hydrogen chain. Except for ADAPT methods, we utilize tensor network algorithms to simulate circuits beyond the scale where brute force linear algebra is possible. For Trotterized circuits especially we utilize the time evolving block decimation (TEBD) algorithm~\cite{Vidal_2003}. For ADAPT-VQE, we used our sparse wavefunction circuit simulator (SWCS)~\cite{Mullinax2025swcs,Mullinax2025adapt}. For computing energy errors~\eqref{eqn:ground-state-energy-estimation-problem}, we used the density matrix renormalization group (DMRG) algorithm to compute the reference ground state energy $E_0$, except for the metaphosphate molecules where coupled cluster with singles, doubles, and triples (CCSDT) was used.

\subsection{Quantum Krylov bounds and estimates} \label{sec:methods-krylov}

\begin{figure}
    \centering
    \includegraphics[width=\linewidth]{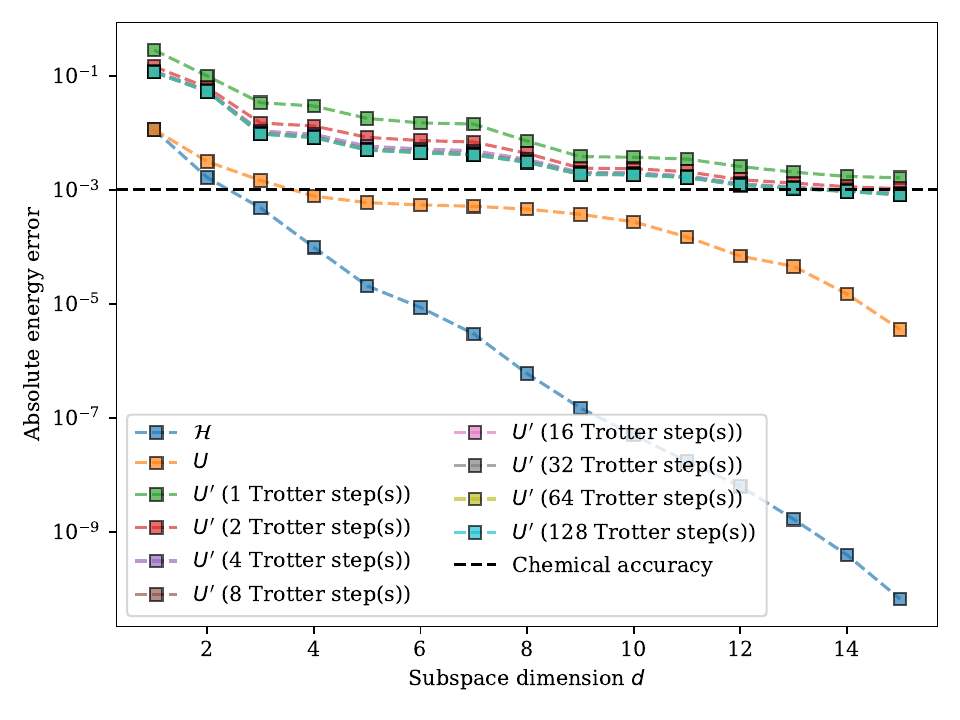}
maf    \caption{Accuracy of quantum Krylov applied to a $n = 6$ qubit linear chain of Hydrogen atoms as a function of the subspace dimension $d$ and the number of Trotter steps. For reference, quantum Krylov with exact time evolution (infinite Trotter steps) is shown. The initial state $|\phi\rangle$ is chosen such that the overlap with the true ground state $|\psi\rangle$ is $|\< \phi | \psi \>|^2 = 0.85$.}
    \label{fig:krylov-convergence-hchain}
\end{figure}

In parallel with the rich literature on classical Krylov convergence and bounds, analogous results have recently been demonstrated for unitary quantum Krylov with $U = \exp(- i \H t)$~\cite{kirby2024analysis,epperly2022theory}. 
In particular, Ref.~\cite{epperly2022theory} shows the relatively simple bound relating the subspace dimension $d$ to the required accuracy $\epsilon$ via
\begin{equation}
    0 \le E - E_0 \le 8 \Delta E_{N - 1} \frac{1 - |\gamma_0|^2}{|\gamma_0|^2} \left( 1 + \frac{\pi \Delta E_1}{\Delta E_{N - 1}} \right)^{-2 d} 
\end{equation}
where $\Delta E_i := E_i - E_0$ is the $i$th energy gap, $N$ is the number of eigenvalues / dimension of the system, and $|\gamma_0|^2$ is the overlap of the initial state with the ground state. Unfortunately this bound is often very loose, e.g. for the six-qubit Hydrogen chain numerically studied in Fig.~\ref{fig:krylov-convergence-hchain}, taking $|\gamma|^2 = 0.975$ and $E - E_0 = \epsilon = 10^{-3}$ yields $d \le 4.85 \cdot 10^{15}$, while numerically we see a much more modest $d \sim 15$ is sufficient to achieve chemical accuracy. 

\begin{figure}
    \centering
    \includegraphics[width=\linewidth]{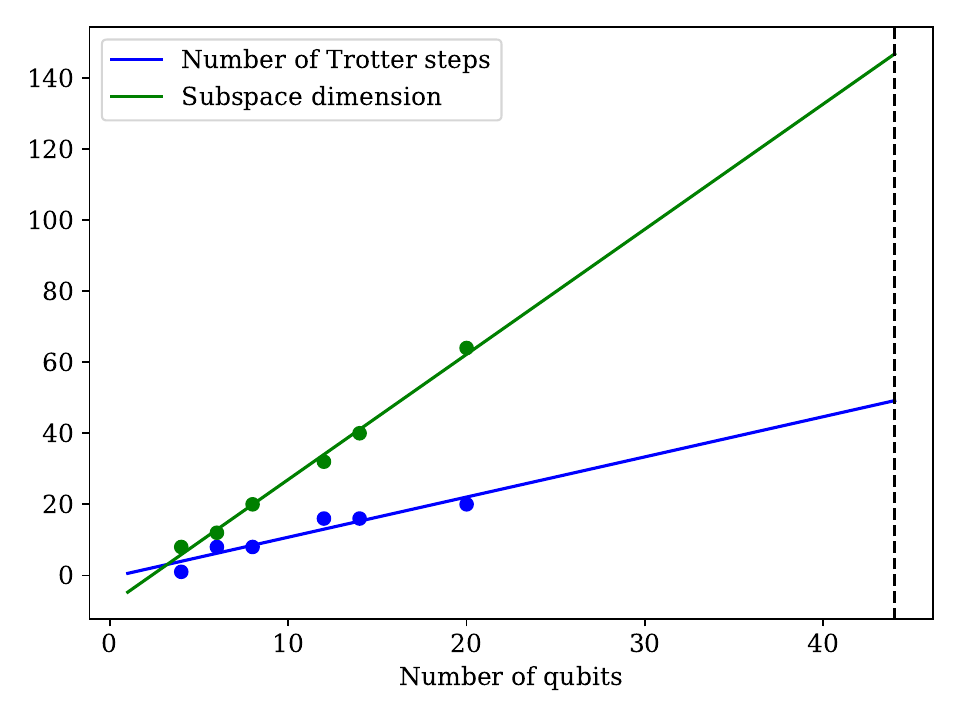}
    \caption{Estimation of the required subspace dimension $d$ and number of Trotter steps $n_T$ for quantum Krylov. Points show values computed by exact simulation to reach chemical accuracy on hydrogen chains of $n_H = n_Q / 2$ atoms. We fit a line to these points to extrapolate to the metaphosphate system size (vertical dashed line), and take these as estimated lower bounds for the more complicated metaphosphate systems.}
    \label{fig:krylov-extrap-estimates}
\end{figure}

Because analytical bounds are often loose, and because we are incapable of directly simulating quantum Krylov for the large, complex metaphosphate systems considered here, we compute resources for smaller, simpler systems and extend trend lines to estimate the subspace dimension $d$ and number of Trotter steps $n_T$ for metaphosphate. This is illustrated in Fig.~\ref{fig:krylov-convergence-hchain} and in Fig.~\ref{fig:krylov-extrap-estimates}. First, Fig.~\ref{fig:krylov-convergence-hchain} shows an example of estimating $d$ and $n_T$ for a Hydrogen chain with $n_H = 3$ atoms. We see from this plot that, while exact unitary evolution reaches chemical accuracy at a very small subspace dimension $d$, Trotterized time evolution requires larger values of $d$, in this case $d \approx 15$. Here, the difference in error is notable but relatively small for different numbers of Trotter steps. By performing the same numerical analsysis for larger values of $n_H$, we obtain the data points and trend lines in Fig.~\ref{fig:krylov-extrap-estimates} which we extrapolate to the size of the metaphosphate systems. Because hydrogen chains are significantly simpler Hamiltonians than the metaphosphate models, we take these to be lower bounds on the resources required for metaphosphate. While these are estimates, we remark that they are significantly more useful than analytical bounds and, in the absence of being able to directly simulate the system, provide a reasonable means for estimating resources.

\subsection{Quantum phase estimation bounds} \label{sec:methods-qpe}

Much work has been done to bound errors on Trotterized quantum phase estimation, especially for molecular Hamiltonians. Here, we follow the perturbative method proposed in~\cite{MartinezMartinez2023assessmentofvarious} where the Trotter-Suzuki expansion of the evolution operator leads to a system evolving under an effective Hamiltonian, the energy of which is close to the original Hamiltonian. Let $\H = \sum_{\mu=1}^L H_{\mu}$ so that the first-order Trotterized evolution operator for the system is $e^{-i H_1 t} e^{-i H_2 t} \cdots e^{-i H_L t}$. For a system with ground state $\ket{\phi_0}$ evolving for time $t$, the first-order correction to the ground-state energy is
\begin{equation}
    \label{eq:qpe-energy-shift}
    E^{(1)}_{\mathrm{GS}} = \bra{\phi_0} V_2 \ket{\phi_0} t^2
\end{equation}
where
\begin{equation*}
    V_2 := \frac{-1}{24} \sum_{\mu=1}^{2M - 1} \sum_{\nu = \mu+1}^{2M}
    \sum_{\nu' = \nu}^{2M} \bigg(1 - \frac{\delta_{\nu,\nu'}}{2}\bigg)
    [H_{\nu'}, [H_{\nu}, H_{\mu}]].
\end{equation*}
For large systems with hundreds of thousands to millions of terms, this energy correction is intractable to evaluate exactly. Instead, we bound this quantity as follows. Let $\H = \sum_i h_i P_i$ be the expansion of $\H$ in the Pauli basis. For any Pauli strings, $|\bra{\phi_0} [P_{\nu'}, [P_{\nu}, P_{\mu}]] \ket{\phi_0}| \leq ||[P_{\nu'}, [P_{\nu}, P_{\mu}]]|| \leq 1$. Let $h_\textrm{max} = \max_{\mu} |h_\mu|$. By the bilinearity of the commutator, we have
\begin{equation}
    \label{eq:qpe-energy-bound}
\begin{split}
    |E_{\mathrm{GS}}^{(1)}| &\leq \frac{1}{24}
    \sum_{\mu=1}^{2M - 1} \sum_{\nu = \mu+1}^{2M}
    \sum_{\nu' = \nu}^{2M} \bigg(1 - \frac{\delta_{\nu,\nu'}}{2}\bigg) |h_{\mu} h_{\nu} h_{\nu'}| \\
    &\leq \frac{1}{24} h_\textrm{max}^3 M^3.
\end{split}
\end{equation}
From the desired accuracy $\epsilon$ and this bound on $E_{\mathrm{GS}}^{(1)}$, the maximum length of a Trotter step can be determined via
\begin{equation}
    \label{eq:qpe-dt-bound}
    \delta t \approx \sqrt{\frac{\epsilon}{|E_{\mathrm{GS}}^{(1)}|}}.
\end{equation}
Instead of the energy $E_0$ of the ground state, quantum phase estimation predicts the phase $E_0 t$, where $t$ is the chosen evolution time. Since phase estimation only predicts this phase modulo $2 \pi$, we adopt the convention in~\cite{Ni2023lowdepthalgorithms} by taking
\begin{equation}
    \label{eq:qpe-evolution-time}
    t = \frac{\pi}{4||H||_2},
\end{equation}
to ensure all eigenvalues are in the interval $[-\pi/4, \pi/4]$. The number of Trotter steps is then $n_T = t / \delta t$.

\subsection{Circuit compilation/compression} \label{sec:methods-compilation}

\begin{figure}
    \centering
    \includegraphics[width=\linewidth]{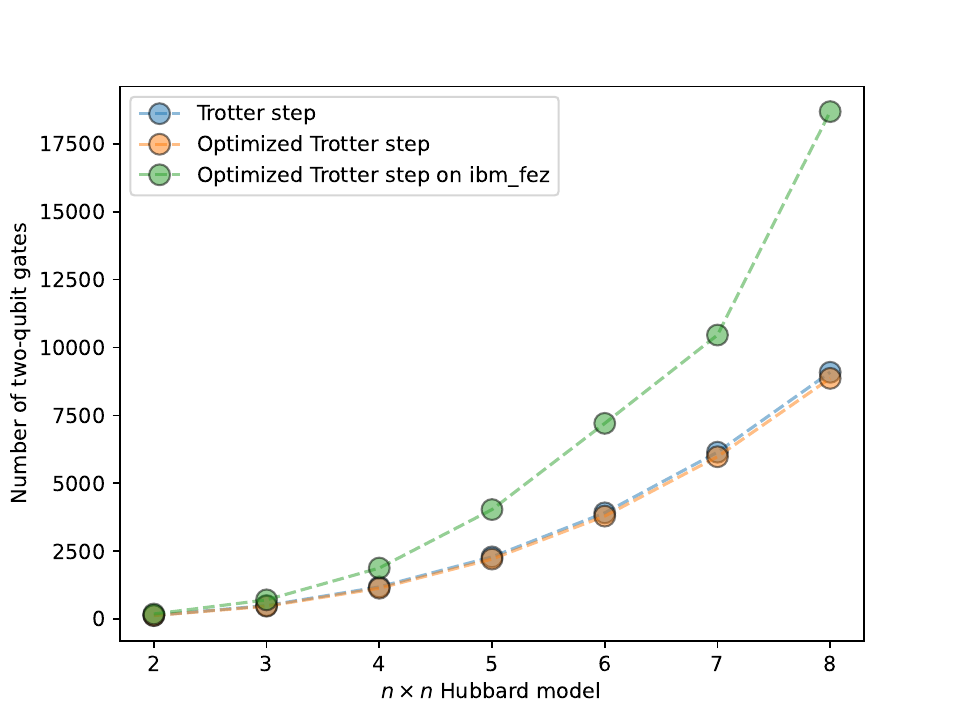}
    \caption{Number of two-qubit gates for a single Trotter step of the $n \times n$ Hubbard model. The three curves show the base number with all-to-all connectivity (blue), the number after circuit optimization/compression with all-to-all connectivity (orange), and the number after circuit optimization/compression with heavy hex connectivity (green), for example on the IBM Fez quantum processor.}
    \label{fig:hubbard-trotter-compilation}
\end{figure}

To illustrate the overhead of particular hardware, we show the difference in two-qubit gates between all-to-all connectivity (assumed in Table~\ref{tab:resources}), which is typically available on neutral atom and trapped ion quantum computers, with more constrained connectivity, typically arising in superconducting qubit quantum computers, in Fig.~\ref{fig:hubbard-trotter-compilation}. Here, we first generate a Trotter step circuit for the $n \times n$ Hubbard model~\eqref{eqn:hubbard} via the Paulihedral method described in Sec.~\ref{sec:methods-time-evolution}. We then optimize this circuit utilizing the Qiskit transpiler, still considering all-to-all connectivity. Then, we compile and optimize the circuit to the ``heavy hex'' topology of IBM quantum computers. While the overhead is negligible for small systems up to around $20$ qubits, we see that for systems around $40$ qubits the number of two-qubit gates doubles. Since the metaphosphate molecule is markedly more complex than the Hubbard model, we expect an overhead of at least two for the gate counts in Table~\ref{tab:resources} compiled to IBM quantum computers.

While this result applies to Trotter circuits, we also studied the impact of state-of-the-art compilation frameworks on the two-qubit gate counts specifically for ADAPT circuits.
For all-to-all connectivity targets, we compared three popular quantum circuit instantiation frameworks with built-in compilation workflows: Lawrence Berkeley National Laboratory's Bqskit, Quantinuum's Tket, and IBM's Qiskit~\cite{younis2021berkeley, Sivarajah_TKET_A_Retargetable_2020, javadi2024quantum}. We used these frameworks to transpile the ADAPT circuits to a basis set composed of single-qubit rotations (U3) and two-qubit CNOT and CZ operations. Then, using Qiskit, we transpiled the circuits obtained in the all-to-all step into the basis set and topology of IBM Fez --- a 156-qubit representative of the heavy hex QPU family. We also compared this approach against compilation to IBM Fez directly, skipping the all-to-all intermediate stage. The results are shown in Table~\ref{tab:compilation_strategy_comparison}. These results showcase the overhead due to different quantum hardware. They also demonstrate that a compilation flow that is best for an all-to-all target is not necessarily the one that yields the lowest two-qubit counts once the circuit is mapped to a heavy-hex device, as seen by ``Bqskit \& Qiskit'' outperforming ``Tket \& Qiskit'' for IBM Fez. Likewise, the compilation strategy that performs best at a given ADAPT iteration depth does not necessarily remain optimal for other depths. Both observations are supported by analogous compilation comparisons done across different operator pool selections for ADAPT-VQE circuits. Taken together, these results suggest that careful, hardware-aware compilation benchmarking is required when estimating and comparing the two-qubit gate costs of near term quantum algorithms.

\begin{table}[h] 
    \centering
    \begin{tabular}{|l|c|c|c|c|c|c|}
        \hline
        & \multicolumn{6}{c|}{$n_{2Q}$} \\
        \cline{2-7}
       Compiler \quad  & \multicolumn{3}{c|}{All-to-all target} & \multicolumn{3}{c|}{Heavy hex target} \\
        \cline{2-7}
         & 1 it. & 10 it. & 20 it. & 1 it. & 10 it. & 20 it. \\
        \hline
        Bqskit (\& Qiskit) & 21 & 298 & 638 & 21 & \textbf{474} & 1351 \\
        Tket (\& Qiskit) & \textbf{11} & \textbf{140} & \textbf{278} & \textbf{18} & 525 & \textbf{1279} \\
        Qiskit & 26 & 280 & 578 & 39 & 695 & 1751 \\
        Qiskit Direct & - & - & - & 39 & 709 & 1695 \\
        \hline
    \end{tabular}
    \caption{Impact of different compilation strategies on two-qubit gate counts ($n_{2Q}$). Each column corresponds to the combination of a target architecture and one of three increasingly complex ADAPT circuits (1, 10, and 20 iterations). For each target-circuit pair, the gate count obtained with each compilation strategy is shown, with the smallest count in bold.}
    \label{tab:compilation_strategy_comparison}
\end{table}

\subsection{Subroutines for energy estimation} \label{sec:methods-energy-estimation-subroutines}

Estimating the energy of a variational state requires evaluating
$\langle \psi(\theta) | H | \psi(\theta) \rangle$ for a Hamiltonian expressed as a sum of arbitrary
operators $H = \sum_g c_g h_g$. The total number of measurements needed
to achieve precision $\epsilon$ is governed by the standard variance expression~\cite{crawford2021si}
\begin{equation} \label{eqn:num-shots}
N_{\mathrm{shots}} \approx \frac{1}{\epsilon^2}
\left(\sum_{g} \sqrt{\mathrm{Var}(h_g)}\right)^2
\end{equation}
where each $h_g$ is a group of terms measured in a single basis. 
Commuting-group approaches sort Pauli operators according to either
qubit-wise commuting~\cite{verteletskyi20qwc} or full commuting~\cite{chong20fc}, and then measure
each group in a shared basis. Practical grouping procedures typically use greedy
sorted insertion~\cite{crawford2021si} strategies, which incrementally build groups to minimize the
measurement variance objective. We use this method to produce $N_\textrm{groups}$ in Table~\ref{tab:resources}.

In addition to the number of groups, we estimate the number of shots~\eqref{eqn:num-shots} required to compute the energy for each metaphosphate system to chemical accuracy (here taken to be $\epsilon = 3 \cdot 10^{-3}$). The results are shown in Tab.~\ref{tab:owp_shot_counts}.
To obtain these results, we approximate each Hamiltonian as a matrix product operator (MPO) with bond dimension $\chi_{\text{MPO}}$, then run DMRG~\cite{Schollwock2011DMRG} with bond dimension $\chi_\text{MPS}$ to obtain an approximate matrix product state (MPS) ground state $|\Psi(\chi_\text{MPS} )\rangle$. For each group $h_g$, we then compute $\text{Var} (h_g) =\sqrt{\sum_i c_i^2\text{Var}(P_i)}$ for $P_i\in h_g$, where $\text{Var}(P_i) = 1 -\langle \Psi(\chi_\text{MPS}) |  P_i | \Psi(\chi_\text{MPS}) \rangle^2$.
We utilize an MPO bond dimension twice that of the desired DMRG bond dimension, and increase this bond dimension until the shot counts converge to a specified threshold. For comparison, we also include an estimate based on the Hartree--Fock (HF) which provides a lower bound on the
required number of shots, as well as an estimate based on the maximally mixed (MM) state
\[
N
=
\frac{1}{\epsilon^2}
\left(
\sum_g
\sqrt{\sum_{P_i \in g} c_i^2}
\right)^2
\]
which provides a a worst-case/upper bound on the required number of shots. While we use the commuting group partitioning method to estimate shots here, we remark that we discuss additional strategies in Sec.~\ref{sec:improvements}.

\begin{table}[t]
\centering

\begin{tabular}{lccc}
\hline\hline
System & $\chi_{\text{MPO}}$ & $\chi_{\text{MPS}}$ & Shots \\
\hline
PO$_3^-\cdots$H$_2$O  & --  & HF   & 47,476,196 \\
 & 100 & 50  & 58,127,824 \\
 & 200 & 100 & 58,365,775 \\
 & 800 & 400 & 58,826,716 \\
 & -- & MM & $44.8\times 10^9$ \\
\hline
H$_2$PO$_4^-$  & --  & HF   & 59,012,325 \\
  & 100 & 50  & 71,657,739 \\
  & 200 & 100 & 71,821,826 \\
  & 800 & 400 & 72,290,143 \\
 & -- & MM & $44.9\times 10^9$ \\
\hline\hline
\end{tabular}
\label{tab:owp_shot_counts}
\caption{Estimated shot counts for computing energies to $3 \cdot 10^{-3}$ Ha. HF indicates the Hartree--Fock state and MM indicates the maximally mixed state. Other rows use the strategy of increasing DMRG bond dimensions until the shot counts converge to a specified threshold as described in the main text.}
\end{table}

\section{Discussion}

\subsection{Limitations}

Our gate counts in Table~\ref{tab:resources} do not include diagonalization circuits for energy estimation (in ADAPT-VQE and quantum Krylov), nor do they include initial state preparation (in quantum Krylov and quantum phase estimation). The cost of diagonalizing a commuting group of $n$-qubit Paulis is known to scale as $O(n^2 / \log n)$~\cite{dalFavero2025measurement}. The cost of state preparation for quantum Krylov and QPE can be non-trivial as these algorithms require significant support with the ground state to converge. Many strategies exist for choosing initial states --- for (bio)chemical Hamiltonians, the Hartree-Fock state is often taken; one can also run (ADAPT-)VQE for a given accuracy smaller than chemical accuracy and use this as the initial state. 

Our resource counts include only quantum and not classical resources. This is especially notable for variational quantum algorithms which are designed to offload as much resources to classical computers as possible, relying on classical optimization algorithms to construct the ansatz. Quantum Krylov methods also utilize classical resources but to a much lesser extent --- namely, to solve the $d \times d$ generalized eigenvalue problem~\eqref{eqn:krylov-generalized-eigenvalue-problem}. Quantum phase estimation utilizes essentially no classical resources (besides circuit compilation/compression which are also present in VQE and quantum Krylov).

\subsection{Improvements} \label{sec:improvements}

It is possible that improved measurement strategies could reduce the total number of shots relative to the commuting group partitioning method we have considered here. For example, recent work has shown that exploiting
fine-grained or partially overlapping commutativity allows substantially fewer
fragments while maintaining Clifford measurement circuits~\cite{dalFavero2025measurement}.
Relaxing commutativity further is possible, where each
fragment is required only to admit a product-of-local-operators diagonalization
with locality at most $k$~\cite{sawaya2024non}. This shifts cost from
measurement into shallow basis-change circuits, enabling substantially fewer
measurement settings at a controlled overhead in depth. Related approaches
include ``ghost''~\cite{choi2022ghost} and ``fluid''~\cite{choi2023fluid}
partitioning methods, which enforce specialized fragment structure while preserving efficient
measurement bases. Randomized measurement schemes such as classical shadows provide another
approach in which many observables can be estimated simultaneously from a shared
pool of randomized measurement data~\cite{huang2022learning}. However, empirical
benchmarking indicates that deterministic grouping strategies exploiting
Hamiltonian structure tend to achieve significantly lower shot counts for
electronic structure problems at practical accuracy targets~\cite{yen23deterministic}. Many methods have been optimized for fermionic systems, e.g.~\cite{huggins2021efficient}, and may improve the resources we determined here. We remark that we make code and data publicly available (see \textit{Code and data availability}) such that these improvements may be carried out.

\subsection{Impact}

In this work we provide quantum resource requirements for a problem of significant biochemical and biological importance. Improving the quantum algorithms considered here, or developing new quantum algorithms, to solve the ground state energy estimation problem for these systems would have important implications in metabolism, cellular signaling, and many other reactions important for biological function. Quantum algorithms are worthwhile to consider both for the natural embedding of such biomolecules in quantum information carriers as well as the difficulty with which state-of-the-art classical methods have had with such problems. On the resource estimation side, our work provides, whenever possible, exact resources and/or numerical estimates significantly tighter than analytical bounds. We also extend typical resource estimation work by considering three state-of-the-art algorithms for NISQ, MegaQuop, and FASQ computing, providing a picture of the necessary quantum resources in each regime. We hope our work serves as a launching point for further improvements in quantum algorithms for biochemistry and further improvements in resource analysis and reduction.

\section*{Code and data availability}

Code to reproduce all data and plots is available at \href{https://github.com/rmlarose/compare-vkp}{https://github.com/rmlarose/compare-vkp}.

\section*{Acknowledgements}

This work is supported by Wellcome Leap as part of the Q4Bio Program.

\bibliographystyle{unsrt}
\bibliography{refs}

\onecolumngrid
\appendix

\section{Performance of quantum Krylov methods with and without thresholding}

\begin{figure}
    \centering
    \includegraphics[width=0.32\linewidth]{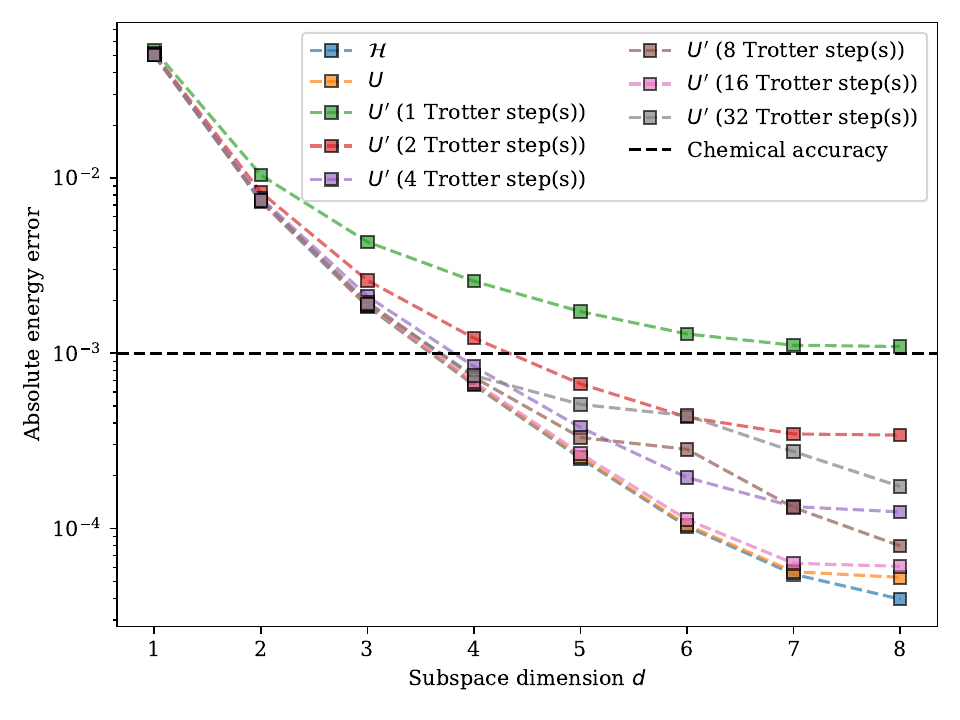}
    \includegraphics[width=0.32\linewidth]{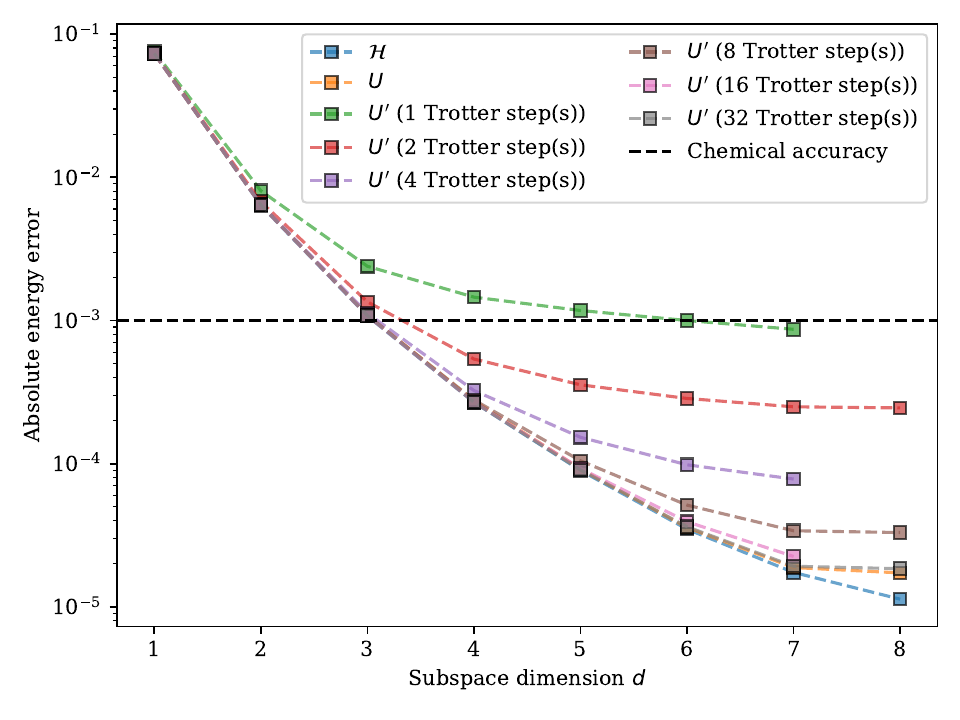}
    \includegraphics[width=0.32\linewidth]{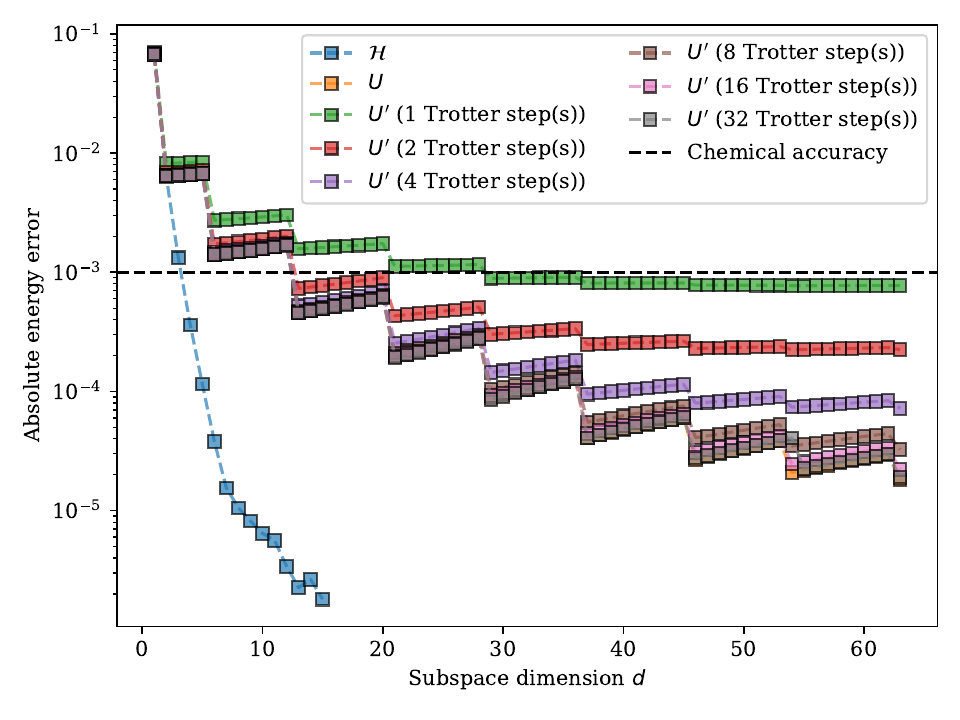}
    \caption{The performance of quantum Krylov for an $n = 8$ qubit Fermi-Hubbard model with no threshold (left), a small negative threshold of $\delta = - 10^{-4}$ (middle), and a small positive threshold of $\delta = 10^{-4}$. The initial state overlap with the ground state for all cases is $0.98$. Thresholding with a (small) positive value improves the stability of the method --- the left two plots only converge for up to $d = 8$, whereas the right plot converges up to $d = 62$. However, a small positive threshold requires a higher subspace dimension to achieve a desired accuracy --- as can be seen in the right plot. This is because thresholding discards the support of small eigenvalues, so if the eigenvalues do not change significantly when the subspace dimension increases, the energy estimate will not change, resulting in the ``step-like'' behavior of the rightmost plot. In the main text, we report the gate counts for quantum Krylov without thresholding since this leads to the fewest resources, although in practical applications thresholding will likely be desired to improve stability.}
    \label{fig:krylov-threshold}
\end{figure}

Thresholding~\cite{kirby2024analysis,epperly22theory} refers to discarding eigenvalues in Krylov matrices~\eqref{eqn:krylov-matrix-elements} when solving the generalized eigenvalue equation~\eqref{eqn:krylov-generalized-eigenvalue-problem}. Nominally, the procedure is designed to improve the stability of quantum Krylov in practice, although it is easy to construct simple examples where thresholding is undesirable~\cite{epperly2022theory}. We illustrate the effect of thresholding in Fig.~\ref{fig:krylov-threshold} for a small Hubbard model. As can be seen, not using a threshold produces lower resource requirements (subspace dimension $d$ and number of Trotter steps $n_T$), at the cost of less stability --- these values of $d \le 8$ were the only ones for which a solution of the generalized eigenvalue could be found, although $1 \le d \le 64$ was considered. Thresholding to a small positive value typically enables one to solve the generalized eigenvalue problem for much larger subspace dimensions, however the support of the subspaces may not change significantly when thresholding because it discards (small) eigenvectors. In practice it will likely be desirable to threshold for stability; however, the resource requirements we report in the main text, unless otherwise noted, do not use thresholding to produce the smallest possible resources for quantum Krylov.

\end{document}